\documentclass[preprint,5p,twocolumn,numbers]{elsarticle}

\usepackage{amssymb}
\usepackage{amsmath}
\usepackage[colorlinks = true,
  linkcolor = blue,
  citecolor = blue,
  urlcolor = blue]{hyperref}
\usepackage{balance}
\usepackage{graphicx}
\usepackage{booktabs}
\usepackage{multirow}
\usepackage{url}

\begin{document}

\title{Risks and Compliance with the EU's Core Cyber Security Legislation}
\author[sdu]{Jukka Ruohonen\corref{cor}}
\ead{juk@mmmi.sdu.dk}
\cortext[cor]{Corresponding author.}
\author[fa]{Jesper L\o{}ffler Nielsen}
\author[sdu]{Jakub Sk\'orczynski}
\address[sdu]{University of Southern Denmark, Denmark}
\address[fa]{Focus Advokater, Denmark}

\begin{abstract}
\textit{Context}: The European Union (EU) has long favored a risk-based approach
to regulation. Such an approach is also used in recent cyber security
legislation enacted in the EU. Risks are also inherently related to compliance
with the new legislation. \textit{Objective}: The paper investigates how risks
are framed in the EU's five core cyber security legislative acts, whether the
framings indicate convergence or divergence between the acts and their risk
concepts, and what qualifying words and terms are used when describing the legal
notions of risks. \textit{Method}: The paper's methodology is based on
qualitative legal interpretation and taxonomy-building.  \textit{Results}: The
five acts have an encompassing coverage of different cyber security risks,
including but not limited to risks related to technical, organizational, and
human security as well as those not originating from man-made actions. Both
technical aspects and assets are used to frame the legal risk notions in many of
the legislative acts. A threat-centric viewpoint is also present in one of the
acts. Notable gaps are related to acceptable risks, non-probabilistic risks, and
residual risks. \textit{Conclusion}: The EU's new cyber security legislation has
significantly extended the risk-based approach to regulations. At the same time,
complexity and compliance burden have increased. With this point in mind, the
paper concludes with a few practical takeaways about means to deal with
compliance and research it.
\end{abstract}

\begin{keyword}
risk analysis, requirements engineering, legal requirements, compliance, conformance, certification
\end{keyword}

\maketitle

\section{Introduction}

For a long time, the EU has favored a risk-based approach to both laws and
lawmaking. In fact, it has been argued that the risk-based governance is of
significant importance and currently is one of the prevailing approaches in the
EU and also elsewhere~\cite{vanderHeijden21}. In Europe, the origins of the
approach trace to the EU's product safety legislation from the circa 1980s
onward~\citep{Ruohonen22ICLR}. The so-called ``new approach'' initiated during
that time later evolved into a decision to consider safety risks when drafting
new laws, negotiating them, and approving them~\citep{EU08}. In essence:
whenever a risk is seen as high, more stringent requirements should be codified
also into a law or multiple laws.

A similar approach has increasingly been used also for the EU's technology
legislation, including but not limited to cyber security laws. When the
Cybersecurity Act (CSA) was approved in 2019, the European Union Agency for
Cybersecurity (ENISA) together with different coordination groups were tasked to
consider cross-border risks, contribute to European and international risk
management standards and frameworks, raise public awareness about cyber security
risks, and develop risk-based cyber security certification
schemes~\citep{EU19}. By now, in 2025, many of these tasks have been either
already completed or are well-underway of being completed. New legislation has
also been enacted.

Yet, according to a recent systematic mapping study, risk-based approaches and
risk analysis itself have received a very limited attention in existing software
and requirements engineering research dealing with
compliance~\cite{Kosenkov25}. This literature review result is surprising to say
the least because of the EU's risk-based paradigm to lawmaking. To this end, the
paper operates within a notable research gap and contributes to filling
it. Furthermore, the risk-based approach has frequently raised questions among
both practitioners and researchers. For instance, the General Data Protection
Regulation (GDPR) has often been considered only in terms of compliance risks
with it~\cite{Li22}, not in terms of risks to natural persons and their
rights. Another related example would be about terminology. For instance,
practitioners have had difficulties to understand what significant harms---and
even fundamental rights---mean in the EU's new artificial intelligence
regulation~\citep{Hanif24}. Such contemplations and uncertainties motivate the
paper's first research question (RQ.1) of interpreting \textit{how different
  risk concepts are framed in the EU's core cyber security laws?} Note that this
RQ.1 already provides a framing for the paper because of the plural form; there
is not just a single risk concept but multiple ones. Subsequently, it is
investigated through interpretation (RQ.2) \textit{whether the framings done in
  the laws indicate convergence or divergence between the laws?}  Finally, the
previous point about significant harms motivates the third and final research
question (RQ.3): \textit{what qualifying adjectives and other words are used in
  relation to the risk concepts in the cyber security laws?}

These RQs place the paper into a debated corner of theoretical papers in
software engineering. As has been recently argued~\citep{Halle24}, conventional
evaluation rubrics and checkbox exercises cannot be used for such papers, which,
out of necessity, require flexibility and leeway in interpreting materials and
building theoretical arguments. Before continuing, it can and should be remarked
that the term checkbox exercises is used intentionally in the previous sentence
because that is what compliance engineering has long been criticized to be
about~\citep{Michalec22, Onose20, Suslov25}. In contrast to such exercises, a
holistic approach has been recommended for interpreting laws, their
requirements, and compliance with them~\citep{Haelterman22, Kosenkov25,
  Ryan23}. Although rigorous definitions are lacking, a notion of holistic
compliance seeks to approach risks by covering multiple legal acts and avoiding
a rigid subject-by-subject viewpoint~\cite{Haelterman22}. In terms of software
engineering, a holistic approach to compliance seeks to cover a software
product's all life cycle stages in relation to the purpose of the given
product~\cite{Kosenkov25}. However, as soon clarified in Subsection~\ref{subsec:
  development and operations}, this characterization cannot be fully satisfied
due to the theoretical nature of the paper. Nevertheless, the holistic approach
suits well to describe also the paper's methodology soon described in the next
Section~\ref{sec: approach}. The analysis of the RQs is presented in
Section~\ref{sec: analysis}. As always in scientific papers, a conclusion and a
discussion ends the paper, in the final Section~\ref{sec: conclusion and
  discussion}.

\section{Approach}\label{sec: approach}

The answers to RQ.1 and RQ.2 are used to build a taxonomy. Although there are
existing taxonomies for cyber security risks~\cite{ReaGuaman17, Waqdan25},
including even taxonomies for taxonomies~\cite{Rabitti24}, to the best of the
authors' knowledge, not much work has been done to relate them to laws, and no
work has been done to relate them particularly to the EU's cyber security
legislation. Regarding taxonomies and their building, an existing guideline was
followed; henceforth, a symbol B denotes the research activities enumerated in
the guideline \cite[the third column in Table 13 on p.~54]{Usman17}. Thus, to
begin with, the goal of the facet-based taxonomy is to improve understanding of
legal notions of cyber security risks, and to help researchers and practitioners
for navigating the risk-based aspects of the EU's five core cyber security laws
(B2 and B4). With this goal in mind, the subject matter (B3) is subsequently
described.

\subsection{Materials}\label{subsec: materials}

The noun \textit{core} in the paper's title is open to an interpretation because
there are numerous cyber security legislative acts in the EU, including many
legislative acts specific to particular industry sectors and particular
technologies~\cite{Ruohonen25Bus}. The legislative act on the operational
resilience for the financial sector~\citep{EU22d}, abbreviated as DORA, would be
a good example of a sector-specific legislative act. Nevertheless, the five laws
enumerated in Table~\ref{tab: laws} can all be reasonably interpreted to
together form the core legal cyber security arsenal of the EU at the time of
writing (B6). As a footnote-style clarifying comment, the legislative acts shown
in the table are ordered according to the presentation order in
Section~\ref{sec: analysis}.

\begin{table}[th!b]
\centering
\caption{The Five EU Laws Covered}
\label{tab: laws}
\begin{tabular}{lll}
\toprule
Law & Abbreviation & Reference \\
\hline
Regulation (EU) 2019/881 & CSA & \cite{EU19} \\
Regulation (EU) 2016/679 & GDPR & \cite{EU16b} \\
Regulation (EU) 2024/2847 & CRA & \cite{EU24a} \\
Directive (EU) 2022/2557 & CER & \cite{EU22b} \\
Directive (EU) 2022/2555 & NIS2 &  \cite{EU22a} \\
\bottomrule
\end{tabular}
\end{table}

Of the legislative acts listed, the CER and NIS2 directives, which are
abbreviations for critical entities resilience and network and information
security, are primarily about critical infrastructure protection, the Cyber
Resilience Act (CRA) concerns cyber security of network-connected products,
including software products, the CSA's main content was already briefly
summarized in the introduction, and the GDPR concerns protection of personal
data related to natural persons. Although the GDPR is oftentimes interpreted as
a privacy law~\cite{Li22, Sokolovska18}, data protection is more than that, as
manifested also by the GDPR's cyber security
requirements~\cite{Ruohonen25ESPREb}. Thus, also the inclusion of the GDPR to
the analysis is well-justified.

% Although all five laws are risk-based, some perhaps more than others, the CRA
%is the only one for which the new approach and the related
%guideline~\cite{EU08} were explicitly followed during policy-making.

\subsection{Methodology}\label{subsec: methodology}

The paper is about law and legal requirements. Although computational law and
legal technologies have made advances, interpretation is still there both as a
cornerstone and as the primary methodology~\cite{Hildebrandt24}. If interpreting
laws, or legal reasoning, is defined to be about studying legal conditions and
legal effects~\cite[pp.~28--30]{Hildebrandt20}, the paper's focus is on the
legal conditions specific to risks. These conditions will be analyzed by a
semantic analysis of selected provisions. Thus, a contextual interpretation,
including case law, is excluded. Interdisciplinarity is also increasingly
shaping particularly research and practice of technology
law~\cite{Guerra18}. The same applies to risk
regulation~\cite{Sibony17}. Interdisciplinarity has increasingly characterized
also cyber security research~\cite{AndersonMoore09}. While legal reasoning
provides the methodology, the framing is thus broader.

If cyber security is seen as a branch of computer science and software security
as a branch of software engineering, which in itself can be seen as a branch of
computer science, the paper operates in the intersection of law, computer
science, and software engineering (B1). With this interdisciplinary approach,
the paper also seeks to contribute to lessening of the many enduring problems
hindering, or sometimes even preventing, interdisciplinary research in software
engineering~\cite{Hyrynsalmi25}. Regarding software engineering, furthermore,
the closest reference point is requirements engineering in general and legal
requirements in particular (B1). The notion of legal requirements guide also
the~methodology.

The research questions frame the legal interpretation towards conceptual
analysis, which is a common methodology both in law and software
engineering. Regarding the latter, the paper can be characterized as operating
in theorization for understanding, using qualitative interpretation,
sense-making, and taxonomy-building (B5) as loose methodological building
blocks~\cite{Ralph19}. The previous point about interdisciplinarity gives a
context for the interpretative sense-making of the laws and conceptual
taxonomy-building around them. Namely, the risk concepts are interpreted against
a conventional terminology used in cyber security risk management and analysis
literature, standards, and frameworks (B9 and B10). To minimize redundancy and
to ensure feasibility (B8), only six dimensions are considered for the
taxonomy. These risk dimensions and their categories are subsequently
elaborated.

\subsubsection{Risks}\label{subsec: risks}

\begin{figure}[th!]
\centering
\includegraphics[width=\linewidth, height=7cm]{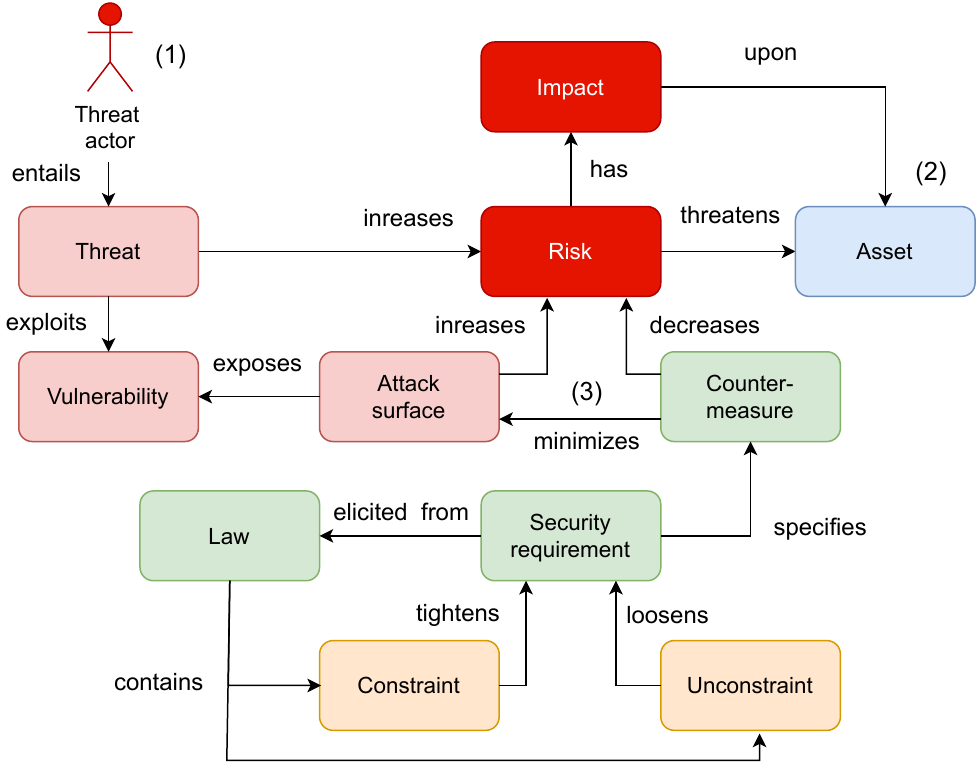}
\caption{The Risk and Requirements Engineering Concepts Used}
\label{fig: concepts}
\end{figure}

Without a substantial divergence from existing frameworks and guidelines, such
as \cite{Alberts16} and \cite{ENISA23}, Fig.~\ref{fig: concepts} elaborates and
summarizes the basic but fundamental concepts used. The figure serves also as a
classifying framework regarding the focal points toward which risk concepts are
framed in the legislative acts. By drawing on closely related ideas presented in
the literature~\cite{Backman23, Tatam21}, the focal points are interpreted to
signify ($1_a$) a \textit{threat-centric}, an ($1_b$)~\textit{asset-centric}, or
a ($1_c$)~\textit{system-centric} viewpoint to the risk concept.

While recognizing that most risks today are man-made~\cite{Wagner16}, including
those related to grand challenges such as the climate change, it is still
sensible to evaluate whether ($2_a$) \textit{only man-made risks} are present or
whether ($2_b$) \textit{other risks too} are considered in the legislation. As
signified already by Fig.~\ref{fig: concepts}, cyber security, like all security
concepts, are about man-made risks because all conventionally considered threat
actor types have distinct intentions to carry out offensive
attacks~\cite{deBruijne17}. Although unintentional accidents happen also in
cyber security~\cite{Carroll14}, they too are man-made, and although cyber
security attacks can cause also safety incidents, intentionality is also
commonly used to distinguish security from safety~\cite{Michalec22}. Thus, in
contrast, a natural disaster affecting important assets is something that can be
seen as belonging to the category of other risks.

In addition, the analysis evaluates whether a ($3_a$)~\textit{probabilistic} or
($3_b$)~\textit{non-probabilistic} understanding of a risk is present. Given the
cyber security context, the former operates with the conventional formula:
``risk = probability of occurrence of a threat $\times$ impact of a
threat''~\cite{Crotty22, ENISA23, Shirey07, Waqdan25}. Despite the noun formula
and due to various problems with quantification~\cite{Michalec22, Vielberth25},
qualitative methods are typically used, often with a help of risk matrices
within which occurrence probabilities denote rows and impacts
columns~\cite{Crotty22}. If a probabilistic understanding is present also in a
law, such matrices seem like a good starting point also in practical risk
analyses.\footnote{~It could be also argued that risk matrices are either
non-probabilistic or only crudely probabilistic. A related point is that there
is no universal definition for a non-probabilistic risk. Examples include a
crossing of a threshold value~\cite{Bernhard21, Vielberth25}, ``the energy
required to change the system state from acceptable to
faulty''~\cite[p.~11]{Mendes23}, and the concept of imprecise
probabilities~\cite{Walley91}. Given these points, a probabilistic understanding
is seen to be present in case the formula, or something closely related to it,
appears in a law. Then, a non-probabilistic understanding is interpreted simply
as an antonym.}

The probabilistic understanding is theoretically important because it entails
also the further concepts of acceptable and residual risks. The former refers to
a ``risk that is understood and tolerated by a system's user, operator, owner,
or accreditor'', whereas the latter is about a ``portion of an original risk or
set of risks that remains after countermeasures have been
applied''~\cite{Shirey07}. Given these definitions, the analysis evaluates
whether ($4_a$) or not ($4_b$) either \textit{acceptable risks} or
\textit{residual risks} are explicitly or implicitly acknowledged in the
legislative acts that fall within the paper's scope.

Regarding the system-centric viewpoint, a noun system is used because
particularly the CRA covers both hardware and software. With this point in mind,
a system-centric viewpoint to risks is seen to operate closer to technical
details, dealing with attack surfaces, technical countermeasures, and more or
less concrete security requirements originating from legislation. This
characterization allows to also distinguish ($5_a$)~\textit{technical security}
from ($5_b$)~\textit{organizational security}, ($5_c$)~\textit{human security},
and ($5_d$)~\textit{national security}. Social engineering and insider threats
are examples of human security, whereas risk analysis itself can be seen as a
core part of organizational security~\cite{Chockalingam23, Suslov25}. These four
security concepts provide additional taxonomy-building material for
answering~to~RQ.1.

Of the concepts shown in Fig.~\ref{fig: concepts}, particularly an asset, which
may be tangible or intangible~\cite{Rabitti24, Ruan17}, must be acknowledged as
a problematic concept; among other things, it is a criticizable concept to
describe natural persons. Although also humans can be seen to have
vulnerabilities~\cite{Rebrean25}, the vulnerability concept refers only to
technical software or hardware vulnerabilities, and any negative impacts upon
natural persons, including any harm done or caused, are conveyed through the
impact dimension. A related point is that the asset as a concept is only
considered in terms of what the legislative acts are primarily seeking to
protect and for whom cyber security is---or should be---designed and
implemented. The GDPR serves well to clarify the point: personal data breaches
have often had significant impacts upon companies, including their intangible
assets such as brands, but natural persons have still almost always been the
``real'' victims~\cite{Ruohonen24IWCC}. Against these backdrops, the concept is
primarily used in order to maintain coherence with the risk management and risk
analysis body of knowledge. A further related point is that the term
countermeasure is interpreted broadly. If the goal is to protect natural
persons, for instance, also a notification about a severe data breach delivered
to them can be interpreted as a countermeasure against identity thefts and other
threats.

\subsubsection{Development and Operations}\label{subsec: development and operations}

Compliance is difficult to deduce about without a context and a real-world
case. Among other things, a distinction between product-oriented and
project-oriented software companies~\cite{Hotomski16} is relevant for compliance
with the EU's cyber security laws. As the latter type of companies develop
software and technologies in general for some particular customers, a compliance
strategy is easier to formulate in a sense that it is known in advance what
needs to be complied with. If a customer operates in a critical sector defined
and referenced in the CER and NIS2 directives, for instance, it is known in
advance that its security requirements for products are more stringent. In
contrast---and as was also pointed out during the CRA's policy
consultation~\cite{Ruohonen25Bus}, product-oriented companies may have
difficulties to conduct risk analyses because a specific context may be
lacking. Such real or potential difficulties also correlate with what was said
about the asset concept in the preceding Section~\ref{subsec: risks}.

Nevertheless, a further taxonomy building block can be drawn analytically:
whether a law's risk concept is about ($6_a$)~\textit{development},
($6_b$)~\textit{operations}, or both. Given \text{DevOps} and related practices,
particularly many software services are about both development-time software
engineering and operating of the services developed. When considering
product-oriented companies, including those developing hardware, the situation
is often different. It is also worth remarking that both organizational and
human security imply that a given risk concept is about operations.

\subsubsection{Convergence and Divergence}\label{subsec: convergence and divergence}

The six taxonomy dimensions are used to evaluate whether there is a convergence
or divergence between the risk concepts used in the legislation covered
(B11). Analogously to mapping analyses commonly used to evaluate standards and
requirements~\cite{Ruohonen25ESPREb}, convergence is \textit{analytically}
understood to simply mean that many, or even most, of the laws map to the same
framings of the underlying risk concepts. As the analysis operates with
theoretical taxonomy dimensions, the mappings are again based on qualitative
interpretation, as has also been the \textit{de facto} method to build
taxonomies in software engineering~\cite{Usman17}. Regarding theoretical
validation (B13), each law is allowed to map to multiple categories in the
dimensions (1), (5), and (6). This allowance of overlaps has sometimes been
discouraged~\cite{Rabitti24}, but it is necessary in the present context because
the categories for the three dimensions do overlap in the laws. Overlaps are
also something with which the holistic approach to compliance is good
at~\cite{Haelterman22}. However, only the categories are allowed to overlap via
the laws.

Although most of the legislative acts reference each other, no explicit mappings
are done between the laws; they only map to each other via the mentioned
categories in the six dimensions. A similar point applies to the categories and
dimensions.

\textit{Theoretically}, the convergence term is interesting because
harmonization was a central goal already in the EU's original ``new approach''
to regulations~\citep{Ruohonen22ICLR}. It is also an explicit goal in the newer
guidelines~\cite{EU08}. Given these goals and broader political objectives,
convergence can be interpreted to represent European integration and
Europeanization in general~\cite[p.~118]{Purnhagen13}. Thus, on one hand, it can
be expected that some convergence should be also present. Particularly the CRA
was strongly motivated by a horizontal harmonization goal, meaning that it is
meant to be a general-purpose cyber security regulation applicable across
multiple different industry sectors. This goal signifies the CRA's close
relation to the rationale used in the EU's product safety legislation. As has
been contemplated in the literature~\cite{Chirico13}, on the other hand, some
divergence is often also a good and desirable goal. For instance, the dimensions
(1) and (5) imply different aspects of cyber security, which may be difficult to
embed into a single unifying legislative act. These points provide a contextual
motivation~for~RQ.2.

\subsubsection{Qualifying Words}\label{subsec: qualifying words}

\begin{figure}[th!]
\centering
\includegraphics[width=\linewidth, height=5cm]{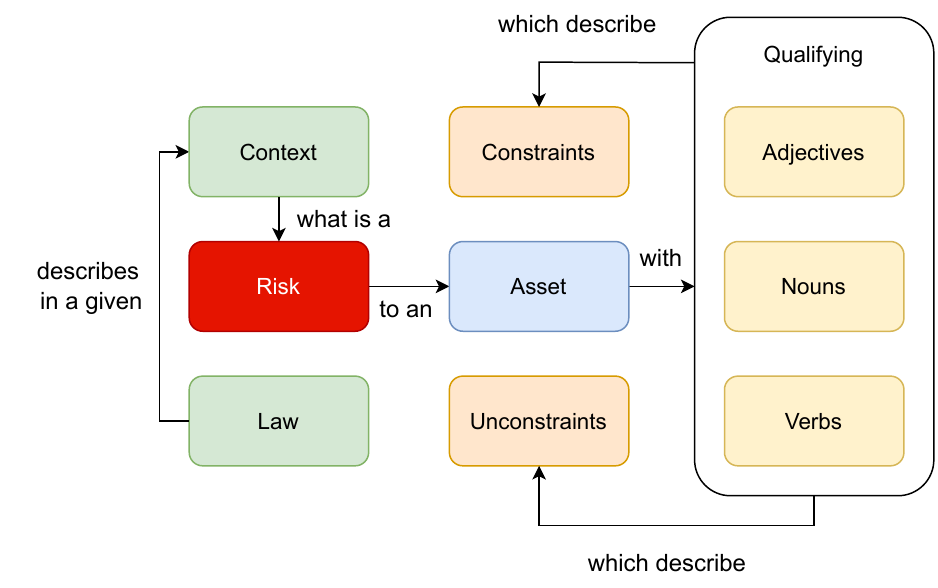}
\caption{Risks and Legal Semantics}
\label{fig: risk semantics}
\end{figure}

With respect to RQ.3, the interest is on so-called qualifying adjectives, such
as the already noted adjective significant for a harm. Although further
semantics are sometimes used, including classifying
adjectives~\cite{Bertoldi07}, the term qualifying adjectives is used without a
particular loss of specificity. Regarding other words, the interest is also on
all nouns and verbs giving a specific meaning to a given term used in the laws
(see Fig.~\ref{fig: risk semantics}). To simplify the terminology and semantics,
hereafter, the noun qualifying is used to refer also to these contextualizing
nouns~and~verbs.

As also seen from Figs.~\ref{fig: concepts} and \ref{fig: risk semantics}, the
qualifying words are interpreted to either tighten or relax a legal requirement
for a cyber security countermeasure. As such, the constraints or unconstraints
emerging from these contextualized semantics are weaker and usually vaguer than
explicit exemptions or other provisions often embedded to laws. Thus, a
so-called keyword trap~\cite{Chirico13} should be kept in mind by avoiding too
strong and definite interpretations. Such a trap also bespeaks against using a
wording about (quantitatively) extracting terms (B7) in the present
context. Nevertheless, particularly constraints are important for requirements
engineering and software development. As cyber security itself can be seen to
constrain software and other systems~\cite{Haley06}, the interest is
particularly on the semantic unconstraints that relax security obligations. When
present, these also give flexibility to a design, an implementation, and
possibly operations too.

\section{Analysis}\label{sec: analysis}

In what follows, the analysis proceeds by considering the CSA, GDPR, CRA, CER,
and NIS2, in the order of listing. As there are hundreds of recitals and
articles to cover, feasibility and brevity are maintained by only considering
notable risk-based aspects related to the five taxonomy dimensions. Thus, it
should be kept in mind that not everything about the laws is covered---nor
should be covered. Before continuing to the analysis, a note about presentation
should be made too: when elaborating the qualifying words identified, they are
shown in italics---and even in case of direct quotations.

\subsection{The CSA}\label{sec: csa}

If the auxiliary risk-based tasks noted in the introduction are excluded, the
CSA's main focus is on the development of European cyber security certification
schemes. These are specified in Article 52(2) to provide three levels of
assurance based on ``the probability and impact of an incident.'' Because the
term asset remains implicit, although it can be interpreted to refer broadly to
something that a user of a product certified has, the concept of impact is
impossible to deduce about. An analogous point has been raised regarding the
EU's artificial intelligence regulation~\cite{Almada25}. Also the probability
concept is only vaguely tied to hypothetical threat actors in the CSA.

As the name indicates, the lowest, ``\textit{basic}'' level provides a basic
assurance level; at minimum, a review of technical documentation is required to
attain a certificate at this assurance level. According to Article 53, the
``\textit{basic}'' level certification may be conducted also through a
self-assessment. The subsequent ``\textit{substantial}'' level takes threat
actors into account, seeking to minimize risks related attacks ``carried out by
actors with \textit{limited skills and resources}'', to quote from
Article~52(6). To attain a certificate at this level, at minimum, also the
absence of publicly known vulnerabilities and security testing must be
sufficiently demonstrated. According to Article~52(7), the ``\textit{high}''
level must further minimize ``\textit{state-of-the-art}'' attacks ``carried out
by actors with \textit{significant skills and resources}''. In addition to the
previously mentioned requirements, a product seeking a certificate at this level
must withstand penetration testing conducted by ``\textit{skilled
attackers}''. All in all, in the certification context, the CSA's framing is
firmly on technical security, development, and the context is post-release
security reviews and security testing, including by third-parties. Given the
certification context, the technical security frame makes the CSA's risk concept
system-centric, although the considerations about attackers render the law's
risk concept threat-centric too.

\subsection{The GDPR}

Risk as a term appears numerous times in the GDPR. In what follows, notable
risk-based aspects of the law are briefly analyzed separately, although they are
all closely related to each other.

\subsubsection{Article~25}

The first is the notion of data protection by design and by default (henceforth,
DPDD). As a guiding principle, the DPDD concept is molded to various parts of
the GDPR, including the principles specified in Article~5 and the responsibility
of a data controlled outlined in Article~24~\cite{Bygrave17, Jasmontaite18}. It
is also the title of Article 25, which contains a risk-based guideline for
complying with the DPDD principle. In particular, ``risks of varying likelihood
and \textit{severity for rights and freedoms} of natural persons'' should be
considered. While keeping the terminological note in Section~\ref{subsec:
  methodology} in mind, natural persons are thus the ``assets'' to be protected
in the parlance of risk analysis literature. To protect natural persons,
``\textit{appropriate technical and organisational} measures'' should be
designed and implemented. Thus, the framing is toward both technical and
organizational security. The qualifying adjective ``\textit{appropriate}'' gives
flexibility for designing and implementing the measures based on a context of
specific activity and technology.

Then, the wording about ``varying likelihoods'' signifies a presence of the
conventional probabilistic understanding of the risk concept. With a qualifying
noun ``\textit{severity}'', the impact should be considered in terms of the
``\textit{rights and freedoms} of natural persons'', including their fundamental
rights.\footnote{~When considering the EU's fundamental rights~\cite{EU16a},
particularly the rights to data protection and respect of privacy life are often
at stake. Given the different philosophical notions of privacy~\cite{Solove02},
also the rights to human dignity and integrity may be violated in some cases. In
some extreme cases it could be argued that even the right to liberty and
security is threatened.} While there is a debate whether the GDPR is a
rights-based regulation or a risk-based regulation, the debate is arguably
questionable as risks should be evaluated in terms of the impacts to natural
person's rights, meaning that the two notions are inherently related to each
other~\cite{Geller16}. The DPDD concept entails also further considerations.

As has been pointed out~\cite{Jasmontaite18}, there are six further qualifying
words present in the phrase about taking into account ``\textit{the state of the
  art}, the \textit{cost} of implementation and the \textit{nature},
\textit{scope}, \textit{context} and \textit{purposes} of processing'' personal
data. Of these, the emphasis on purposes is relevant because it signifies the
GDPR's important purpose limitation requirement, which has been discussed in the
literature to a great extent~\cite{Faisal23, Ruohonen23DS}. Like in the EU's
artificial intelligence regulation, there is thus a probabilistic association
between a purpose and a risk~\cite{Almada25}. Regarding the other qualifying
words, particularly ``\textit{nature}'', ``\textit{scope}'', and
``\textit{context}'' pinpoint toward results from risk analyses, which belong to
a concept of impact assessments in the GDPR's
terminology~\cite{Bygrave17}. Although flexibility was noted earlier regarding
countermeasures, the wording about ``\textit{the state of the art}'' , even when
coupled with the noun ``\textit{cost}'', puts a constraint upon designing and
implementing countermeasures; they should not be outdated or otherwise insecure.

\subsubsection{Article~32}

The GDPR's Article~32 specifies risk-based legal requirements to ensure the
security of processing personal data. As the wordings in Article~32 are very
similar to those in Article~25, the security requirements implied by the former
article largely recapitulate what was said in the preceding section. There is
also a debate in the literature whether the two articles can be even separated
from each other~\cite{Jasmontaite18}. In any case, as the security requirements
have already been addressed in the literature~\cite{Ruohonen25ESPREb}, it
suffices to summarize these by remarking that the fundamental confidentiality,
integrity, and availability (CIA) triad should be guaranteed---though, again
with some qualifying terms, including but not limited to ``\textit{the state of
  the art}''. Because authorization, encryption, pseudonymization, and testing
are explicitly mentioned in Article~32, the GDPR can be interpreted as an
asset-centric and a system-centric law dealing with both technical and
organizational security. Hence, it is also about both development and
operations.

\subsubsection{Article 33 and Article 34}\label{subsec: gdpr art 33 and 34}

Data breach reporting obligations are described in Articles~33 and 34. The
former lays down the requirements to notify data protection authorities about
personal data breaches. Regarding qualifying words, the requirement to notify an
authority within 72 hours after having become aware of a breach is accompanied
with a clause: ``unless the personal data breach is unlikely to result in a risk
to the \textit{rights and freedoms} of natural persons''. Thus, a notification
is required only in case there actually is an impact, which is again qualified
with the two nouns ``\textit{rights and freedoms}''. Regarding the impact
concept itself, the data breach notification requirements could be interpreted
against causal risk models; the initial impact would denote a compromise of
personal data assets and the subsequent impact would correlate with a harm to
natural \text{persons~\cite[cf.][]{Woods16}}. In any case, Article~34 further
requires to notify also the natural persons affected but only when there is ``a
\textit{high} risk to the \textit{rights and freedoms}'' of them.

\subsubsection{Article 35 and Article 36}\label{subsec: impact assessments}

Articles 35 and 36 are about data protection impact assessments. Given that the
GDPR does not set out specific requirements on how such assessments should be
done and documented, several options have been used and mentioned in the
literature, among them scenario analyses and quantitative or qualitative risk
analyses, including risk matrices~\cite{Kloza21, Mollaeefar23, Ruohonen25ESPREb,
  Wairmu24}. Then, Article~35 states that an assessment must be done in
particular when ``using \textit{new technologies}'' but only insofar there is a
``a \textit{high} risk to the \textit{rights and freedoms} of natural persons''
when ``the \textit{nature}, \textit{scope}, \textit{context} and
\textit{purposes}'' of processing are taken into account.

With the exception of ``\textit{new technologies}'', the qualifying words are
among those appearing in the DPDD semantics. When a high risk is identified,
Article~35(7)(d) requires to document the ``the measures envisaged to address
the risks, including safeguards, security measures and mechanisms to ensure the
protection of personal data and to \textit{demonstrate} compliance''. Although
multiple words in this phrase could be seen as qualifying, the verb
``\textit{demonstrate}'' can be interpreted to represent all of them because it
signifies what compliance is also about. In other words, the measures designed
and implemented together with documentation of these should prove that at least
negligence is not present. A somewhat similar rationale appears in the EU's
product liability law~\cite{Ruohonen22ICLR}. In any case, when a ``\textit{high}
risk'' is identified in an impact assessment, prior consultation must be sought
from a data protection authority according to Article~36. If risks were
``\textit{insufficiently} identified or mitigated'' according to the assessment,
the authority should then ``provide written \textit{advice}''. The qualifying
words in italics are used to emphasize the previous point about demonstrating
compliance (or a lack of negligence). For that task, there is also help
available from data protection authorities.

\subsection{The CRA}

The CRA is pronouncedly a risk-based law. Together with a notion of criticality,
Article~7(2) uses the concept of risk to justify a classification of
network-connected products into three groups: ``normal'', important, and
critical. A compliance burden increases alongside this hierarchy, as should also
the cyber security guarantees for users. As with the previous analysis of the
GDPR, in what follows, the focus is on articles relevant to software
engineering. It should be further noted that the CRA's alignment toward the EU's
product safety legislation and the artificial intelligence regulation, including
high-risk systems specified therein, as specified in the CRA's Articles~11 and
12, is omitted on purpose and without major consequences for regarding the
RQs.\footnote{~In addition, many lesser risk-based details are excluded. The
examples include risk considerations with CE markings or other visual labels
used to declare conformity (Article~30(3)), risk-based joint activities by
market surveillance authorities (Article 59(1)), and an obligation of so-called
open source software stewards, including large open source software foundations,
to collaborate with market surveillance for mitigating risks~(Article~24(2)).}

\subsubsection{Article~3}\label{subsec: cra art 3}

The CRA provides explicit definitions for risks in paragraphs (37) and (38) of
Article~3. The former paragraph defines a cyber security risk in conventional
probabilistic terms; a likelihood of an incident should be evaluated in
conjunction with an impact characterized through a ``magnitude'' of a ``loss or
disruption''. The second paragraph provides also a definition for a significant
risk, which is characterized to originate from ``\textit{technical
  characteristics}'' and which causes ``\textit{severe} negative impact,
including by causing \textit{considerable} material or non-material loss or
disruption''. Although the qualifying adjectives remain undefined and thus open
to interpretation, already the notion of ``\textit{technical characteristics}''
frame the CRA's risk concept toward the system-centric viewpoint. This viewpoint
becomes clearer when delving a little deeper into the law.

% Despite this characterization, the CRA's risk concept cannot be perceived to be asset-centric because the asset concept itself remains undefined and vague. This point recapitulates what was said about the CSA in Subsection~\ref{sec: csa}. In any case,

\subsubsection{Article~13}\label{subsec: cra art 13}

According to Article~13, manufacturers are obliged to conduct risk analyses
about their products. As specified in the paragraph (22) of the article, market
surveillance authorities may also request risk analysis documents for evaluating
compliance. As with the GDPR's impact assessments (see Subsection~\ref{subsec:
  impact assessments}), the qualifying verb ``\textit{demonstrate}'' also
appears in the paragraph. Such a demonstration applies particularly to the CRA's
explicit requirements. That is, with a few exemptions enumerated in Article~2,
the CRA defines a set of essential cyber security requirements that practically
all products with a networking functionality must implement. These also overlap
with the GDPR's security requirements for products processing personal
data~\cite{Ruohonen25ESPREb}. Another point is that residual (and acceptable)
risks remain unacknowledged in the CRA with respect to its essential
requirements~\cite{Ruohonen25ESPREa}, as they remain also in the GDPR and the
CSA's certification scheme.

A risk analysis must evaluate step by step whether the essential requirements are
``\textit{applicable} to the relevant product''. If they are applicable, it must
be evaluated ``how those requirements are implemented''. To help at the task, a
risk analysis should consider a product's ``\textit{intended purpose}'' and its
``\textit{reasonably foreseeable use}, as well as the \textit{conditions of
  use}'', including the ``\textit{operational environment} or the
\textit{assets} to be protected, taking into account the \textit{length of time}
the product is expected to be in use''. The wording about a ``\textit{length of
  time}'' means that a risk analysis should address a product's whole support
period. The other qualifying words mean that product-oriented companies (see
Subsection~\ref{subsec: development and operations}) should consider also the
typical or ``\textit{foreseeable use}'' of their products when doing risk
\text{analyses---even} when there is no explicit knowledge about who uses a
product, where, how, and why. The earlier point about scenario analyses in
Subsection~\ref{subsec: impact assessments} may help at these risk analysis
considerations.

\subsubsection{Articles 14, 15, 16, 17, 19, and 20}\label{subsec: cra art 14, 15, 16, 17, 19, and 20}

There are six articles that imply different risk-based reporting obligations for
various parties. When considering only the paragraphs in which a risk concept
appears explicitly, the following summary conveys the main points:
\begin{enumerate}
\item{According to Article~15(1), manufacturers as well as other natural or
  legal persons may voluntarily report vulnerabilities affecting a product's
  risk profile either to a national authoritative computer security incident
  response team (CSIRT) or ENISA.}
\item{According to Article~16, a common reporting platform is established. When
  there is ``an \textit{imminent high}'' risk according to a manufacture, ENISA
  will evaluate whether there is also ``a \textit{systemic} risk'' to the whole
  internal market of the EU. If so, ENISA will nothing other national
  authoritative CSIRTs. Regarding the qualifying words, ``\textit{imminent}'',
  ``\textit{high}'', and ``\textit{systemic}'' are not actually defined in the
  regulation, including in Article~3. Thus, the future will tell how the
  interpretations shall be done in practice, and whether standards will play a
  significant role in this aspect.}
\item{According to Article~17(3), ENISA will prepare a technical report every
  other year about risks in products based on the notifications received.}

\item{When either an importer or distributor of products recognizes a
  significant risk, including due to a vulnerability, it must notify the
  corresponding manufacturer and market surveillance authority according to
  Articles~19 and 20.}

\end{enumerate}

However, it is Article~14 that is particularly relevant with respect to
reporting, although and oddly enough, the risk concept does not explicitly
appear in it. It specifies mandatory reporting of ``\textit{actively exploited}
vulnerabilities and ``\textit{severe} incidents''. Definitions for these are
given in Article~3(42) and Article~14(5), respectively. While the definitions
and their qualifying words are interesting, a further discussion can be omitted
on the grounds that these have already been discussed at
length~\cite{Ruohonen25COSE, Ruohonen25IFIPSEC}. For the present purposes, it
suffices to conclude that the various reporting obligations further signify that
the CRA is about both technical and organizational security. The reporting focus
on incidents and vulnerabilities further reinforce the CRA's system-centric
viewpoint to risks, although some of the auxiliary tasks, including ENISA's
evaluations, exemplify also a consideration of assets. In this regard, as
emphasized throughout the CRA's recitals, the grand ``assets'' to be protected
are the EU's internal market, including particularly consumers, and the
twenty-seven EU member states themselves.

\subsubsection{Article~54 and Article~56}

There are two lengthy articles specifying various obligations and procedures for
dealing with significant cyber security risks. When a significant risk is
detected by a market surveillance authority, the authority, possibly in
collaboration with a relevant CSIRT, should conduct a thorough evaluation of the
given product according to Article~54(1). According to Article~54(2), also
``\textit{non-technical} risk factors'' should be taken into account in an
evaluation particularly regarding supply-chain security issues. When such
factors are present, further information and coordination should be done with
the authorities specified in the NIS2 directive. For mitigating significant
risks, a market surveillance authority should use corrective actions against a
manufacturer for mitigating the significant risk. In extreme cases a given
market surveillance authority may even withdraw or recall a product placed on
the internal market. As with the EU's product safety
legislation~\cite{Ruohonen22ICLR}, withdrawals and recalls may be a stronger
deterrent for manufacturers to comply than the financial sanctions specified in
Article~64.

Furthermore, Article~56 specifies that similar market surveillance actions are
done at the EU-level by the Commission in cooperation with ENISA. These points,
including about ``\textit{non-technical} risk factors'' and EU-level
coordination, further emphasize the asset-centric focal point in addition to the
more technical system-centric focal point. Finally, together the risk-based
essential requirements and reporting obligations signify that the CRA is both
about development and operations---though again keeping the points from
Subsections~\ref{subsec: development and operations} and \ref{subsec: cra art
  13} in mind.

\subsection{The CER Directive}

The CER and NIS2 directives too are pronouncedly risk-based. The concept of
risk, as defined in the directives' Article~2(6) and Article~6(9), respectively,
is the same used in the CRA~(see Subsection~\ref{subsec: cra art 3}). Thus, a
probabilistic understanding is again present. That said, the CER's framing of
risks extends well-beyond cyber security.

\subsubsection{Article 6 and Article 7}\label{subsec: cer art 6 and 7}

The authorities of each member state are obliged to conduct comprehensive risk
analyses. According to Article 6(1), both ``\textit{natural} and
\textit{man-made} risks'' must be assessed, ``including those of a
\textit{cross-sectoral} or \textit{cross-border} nature, \textit{accidents},
\textit{natural disasters}, public \textit{health emergencies} and
\textit{hybrid threats} or other \textit{antagonistic threats}, including
\textit{terrorist offences}''. This quotation frames the CER toward national
security. Although various threats are mentioned, however, none of them pinpoint
toward some particular threat actors, and, hence, a threat-centric viewpoint is
a poor choice for describing the directive. The inclusion of nature-based and
healthcare risks mean that the CER directive extends beyond security too.

Closer to cyber security, the national authorities must also identify the
so-called critical entities in their jurisdictions according to Article 6. These
are those who provide ``\textit{essential} services'' or operate critical
infrastructures. Article 7(1) instructs the authorities to use a concept of
``\textit{significant} disruptive effect''. The article lists six criteria for
evaluating such effects. The examples include users relying on an entity's
service, an impact's duration upon economic and societal activities, and an
entity's market share. As the concept of critical infrastructure is explicitly
tied to the concept of asset in Article~2(4), the CER directive relies on an
asset-centric viewpoint with its probabilistic understanding of risks.

\subsubsection{Articles 12 and 13}\label{subsec: cer art 12 and 13}

Once identified, the critical entities must then themselves do comprehensive
risk analyses. According to Article~12(2), these risk assessments must address
the same aspects as the risk assessments done by the public authorities. A
particularly noteworthy aspect of the article is a requirement to evaluate
whether and to which extent a critical entity depends on other critical entities
or sectors and the other way around. Given the earlier point about risk analyses
themselves belonging primarily to organizational security~(see
Subsection~\ref{subsec: risks}), the CER directive can be interpreted to be
about organizational security too.

The CER directive can be seen to be primarily about operating of critical
infrastructures by critical entities. Though, Article~13(1) mandates the member
states to ``ensure that critical entities take \textit{appropriate} and
\textit{proportionate} \textit{technical}, security and \textit{organisational}
measures to ensure their resilience''. Despite the qualifying adjective
``\textit{technical}'', the CER directive does not explicitly state what kind of
technical resilience measures should be implemented and how. Unlike with the CRA
and GDPR~\cite{Ruohonen25ESPREb}, it remains impossible to elicit any actual
requirements from the article. Thus, the case also reiterates the notion of
keyword traps noted earlier in Subsection~\ref{subsec: qualifying words}. That
said, things may change in the future as the Commission is empowered to develop
``non-binding guidelines'' and ``\textit{technical and methodological
  specifications}'' according to paragraphs (5) and (6) in Article~13.

\begin{figure}[th!b]
\centering
\includegraphics[width=\linewidth, height=4cm]{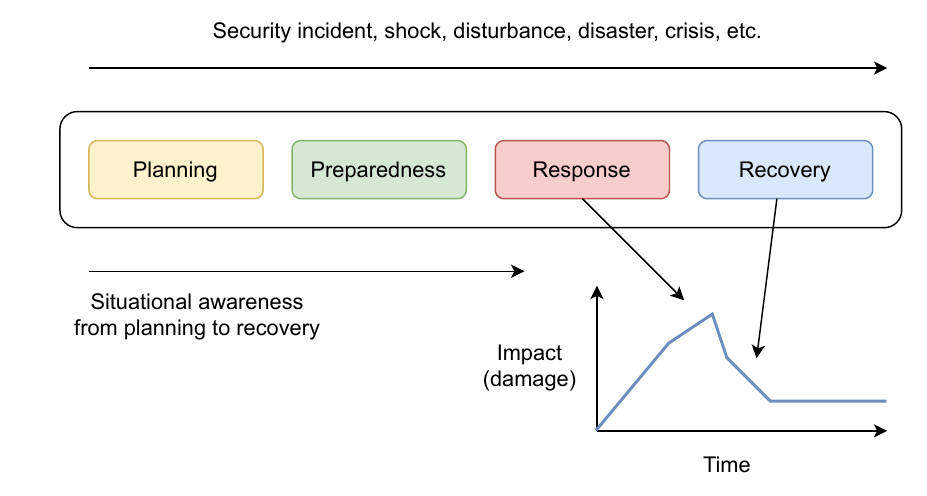}
\caption{A Fairly Conventional Elaboration of Resilience}
\label{fig: resilience}
\end{figure}

Although the term resilience appears in the CRA's name itself, the CER
directive's rationale and goals are closer to it. The concept is defined in
Article~2(2) to mean the ``ability to \textit{prevent}, \textit{protect}
against, \textit{respond} to, \textit{resist}, \textit{mitigate},
\textit{absorb}, \textit{accommodate} and \textit{recover} from an
incident''. Many of the qualifying verbs frame the definition toward incident
management in the cyber security literature. More broadly, the resilience
concept exemplifies a broader trend of not only considering \textit{ex ante}
risks but also responses to risks in legislation~\cite{vanderHeijden21}. By
using the further concepts of planning, preparedness, response, and
recovery~\cite{ENISA24}, an analytical illustration of the resilience concept
can be presented in the form of Fig.~\ref{fig: resilience}.\footnote{~In the
resilience literature the term absorption is often used to emphasize that an
entity should have a capacity to absorb an impact and recover to a normal
state~\cite{Labaka16}. Although there are some hints in related literature that
the response phase has gathered most of the attention~\cite{Kalatpour17}, a
rapid recovery requires planning, which, in turn, increases a level of
preparedness. Planning is necessary also for the detection and analysis of
incidents.}

\subsubsection{Article 14}

Finally, Article 14 obliges the member states to establish frameworks for doing
background checks of natural persons holding sensitive roles in the critical
entities identified. Such checks reiterate the framing toward national
security. These are also an example about human security.

\subsection{The NIS2 Directive}

The NIS2 directive is the CER directive's close cousin and the other way
around. Therefore, in what follows, the analysis focuses only on those
risk-specific aspects that do not have a clear overlap with the CER directive.

\subsubsection{Article 13 and Article 27}

Among the non-overlapping characteristics are national cyber security strategies
the member states are obliged to develop. According to Article~27, these
strategies should ``identify \textit{relevant assets}'' and risks to these,
synchronize risk management procedures with the CER directive's obligations, and
promote a development of risk management frameworks and countermeasures, among
other things. Already the qualifying noun ``\textit{asset}'' reiterates what was
said about the CRA and CER earlier in Subsections~\ref{subsec: cra art 14, 15,
  16, 17, 19, and 20} and \ref{subsec: cer art 6 and 7}. That is, also the NIS2
directive's viewpoint to risks is asset-centric. The point about synchronization
between the CER and NIS2 directives appears also in Article 13 on the
cooperation at the national level. However and despite the synchronization
objective, NIS2 focuses only on man-made cyber security risks. To this end,
recital 72 clarifies that other risks are considered in other laws and handled
through other governance mechanisms.

\subsubsection{Articles 11, 12, and 13}

The NIS2 directive strengthens the role of national authoritative CSIRTs and
increases their obligations. Regarding risks specifically, they are instructed
to also collect and analyze ``\textit{forensic data}'' and provide
``\textit{dynamic} risk and incident analysis''. Although no details for
specific requirements are mentioned, the earlier points about the CER directive
in Subsection~\ref{subsec: cer art 12 and 13} are not a good fit to describe the
NIS2 directive. The reason is that the NIS2 directive have already involved
concrete technical development; the European vulnerability database mandated by
Article~12 is already online~\cite{ENISA25}. With this point in mind, NIS2 can
be interpreted to be also about technical security, having also a system-centric
viewpoint to risks. It is also about both development and operations. The
directive's many requirements for vulnerability coordination and incident
management signify the operational side.

\subsubsection{Articles 15, 16, 17, and 22}

The NIS2 directive further strengthens the European and international network
built around authoritative CSIRTs. It also establishes a new coordination group
as well as a specific group for cyber security crisis management. As specified
in Article~22, the EU-level measures include also risk assessments of critical
supply-chains. Given that the cross-border and crisis management (as opposed to
incident management) aspects have already been discussed in previous
research~\cite{Ruohonen25COSE}, it suffices to summarize them by emphasizing
that they too are strongly risk-based.

\subsubsection{Article 21}

Analogously to the CER directive, so-called essential and important entities, as
defined in the NIS2 directive's Article 3, must conduct comprehensive risk
analyses. Article 21(2) defines ten aspects that must be taken into account
``\textit{at minimum}''. To underline the comprehensiveness, the article uses a
concept of ``an \textit{all-hazards} approach''. Such an approach is usually
understood to involve multiple risks that may be related to each other through
different cascading effects~\cite{Sarmah24}. Thus, they also align with the CER
directive's requirement to consider risk dependencies (see
Subsection~\ref{subsec: cer art 12 and 13}). Interactions between risks and
cascading impacts further reiterate the earlier point about causal risk models
made in Subsection~\ref{subsec: gdpr art 33 and 34}.

As for the ten aspects, these include everything from training and ``cyber
hygiene'' practices to supply-chain security and incident handling. The
technical security focus on the system-centric viewpoint are also present
because cryptography and encryption, multi-factor authentication, continuous
authentication, backups, and other technical aspects are explicitly mentioned
too. In addition to training and related measures, human security is also
explicitly mentioned, and thus the NIS2 directive can be interpreted to be about
technical, organizational, and human security as well. The organizational
security focus is further reinforced by the requirement to consider business
continuity in an ``\textit{all-hazards}'' risk analysis. Given that business
continuity is closely related to the resilience concept~\cite{Meagher24}, the
planning, preparedness, response, and recovery concepts in Fig.~\ref{fig:
  resilience} apply also the NIS2 directive. The incident management
requirements make the relation also explicit.

\subsubsection{Article 32 and Article 33}

Regarding enforcement, the essential and important entities are subject to many
supervisory tasks by public authorities. As specified in Articles~32 and 33,
these include also evaluations of the risks analyses done by the
entities. Furthermore, ``\textit{on-site inspections} and off-site supervision''
are mentioned together with ``\textit{regular and targeted security audits}''
and ``security \textit{scans} based on \textit{objective},
\textit{non-discriminatory}, \textit{fair} and \textit{transparent} risk
assessment criteria''. Again, many of the qualifying words, such as
non-discrimination, are not defined in the directive itself, but can be still
interpreted to represent unconstraints that affect the many supervisory
tasks. Particularly security audits and scans again signify also the technical
security focus and a system-centric viewpoint to risks.

\section{Conclusion and Discussion}\label{sec: conclusion and discussion}

The paper analyzed different risk notions and their framings in the EU's five
core cyber security legislative acts at the time of writing. The goal was to
build a taxonomy by answering to the three RQs specified. The taxonomy developed
through qualitative interpretation is summarized in Table~\ref{tab:
  taxonomy}. In what follows, the taxonomy is first briefly summarized by
considering the answers reached. Afterwards, a few takeaways are discussed for
practitioners and research.

\begin{table*}[t!]
\centering
\caption{A Taxonomy on the Framing of Risks in the EU's Five Core Cyber Security Acts}
\label{tab: taxonomy}
\begin{tabular}{lrlrcccccrr}
\toprule
&& && \multicolumn{5}{c}{Law} \\
\cmidrule{5-9}
&& Category && CSA & GDPR & CRA & CER & NIS2 && $\sum$ \\
\hline
Dimension (1) & ($1_a$) & Threat-centric && $\checkmark$ &&&&&& 1 \\
& ($1_b$) & Asset-centric && & $\checkmark$ & $\checkmark$ & $\checkmark$ & $\checkmark$ && 4 \\
& ($1_c$) & System-centric && $\checkmark$ & $\checkmark$ & $\checkmark$ & & $\checkmark$ && 4 \\
\cmidrule{2-9}
Dimension (2) & ($2_a$) & Only man-made risks && $\checkmark$ & $\checkmark$ & $\checkmark$ & &$\checkmark$ && 4 \\
& ($2_b$) & Other risks too && & & & $\checkmark$ & && 1 \\
\cmidrule{2-9}
Dimension (3) & ($3_a$) & Probabilistic && $\checkmark$ & $\checkmark$ & $\checkmark$ & $\checkmark$ & $\checkmark$ && 5 \\
& ($3_b$) & Non-probabilistic && & & & & && 0 \\
\cmidrule{2-9}
Dimension (4) & ($4_a$) & Acceptable or residual risks && & & & & && 0 \\
& ($4_b$) & No acceptable or residual risks && $\checkmark$ & $\checkmark$ & $\checkmark$ & $\checkmark$ & $\checkmark$ && 5 \\
\cmidrule{2-9}
Dimension (5) & ($5_a$) & Technical security && $\checkmark$ & $\checkmark$ & $\checkmark$ & & $\checkmark$ && 4 \\
& ($5_b$) & Organizational security && & $\checkmark$ & $\checkmark$ & $\checkmark$ & $\checkmark$ && 4 \\
& ($5_c$) & Human security && & & & $\checkmark$ & $\checkmark$ && 2 \\
& ($5_d$) & National security && & & & $\checkmark$ & && 1 \\
\cmidrule{2-9}
Dimension (6) & ($6_a$) & Development && $\checkmark$ & $\checkmark$ & $\checkmark$ & & $\checkmark$ && 4 \\
& ($6_b$) & Operations && & $\checkmark$ & $\checkmark$ & $\checkmark$ & $\checkmark$ && 4 \\
\cmidrule{2-9}
$\sum$ &&&& 7 & 9 & 9 & 8 & 10 && -- \\
\bottomrule
\end{tabular}
\end{table*}

\subsection{Answers and the Taxonomy}

Regarding RQ.1, the framings done in the laws cover all four security concepts
considered. If technical security is present, a viewpoint to risks is also
system-centric though not only system-centric. All laws except the CSA can be
seen to have also an asset-centric viewpoint to risks. As also human security,
organizational security, and national security are present, an overall
convergence is present~(RQ.2). In other words, together the five laws have a
highly encompassing coverage of different aspects of cyber security risks. At
the same time, complexity and compliance burden have increased. Regarding
divergence, the laws cluster into three broad areas with respect to what is
being protected: natural persons (the GDPR), products and their users (the CSA
and CRA), and critical infrastructures (the CER and NIS2 directives). As seen
from Table~\ref{tab: taxonomy}, there are also two notable gaps. Because these
are about risk aspects none of the laws have addressed, the gaps themselves can
be also interpreted as convergence.

% REFACTOR:
As was emphasized in Subsection~\ref{subsec: risks}, the concept of asset is
problematic and requires flexibility in interpretation. Thus, the ``assets''
considered in the laws correlate with the three protection areas; these range
from natural persons and their rights (the GDPR) through the EU's internal
market and users of products (the CSA and CRA) to critical infrastructures and
essential services of European societies (the CER and NIS2
directives).

The CSA is also otherwise an exception because it couples a system-centric
viewpoint to risks with a threat-centric viewpoint. That is, the CSA's
certification scheme involves rigorous technical security testing from a
viewpoint of attackers. This threat-centric viewpoint is also related to
residual and acceptable risks, which none of the laws mention or otherwise
discuss explicitly.  As has been contemplated previously~\cite{deBruijne17},
different threat actors, including their skills and resources, could be seen as
residual risks because threat-centric risk analyses are difficult to do. Given
the certification context, the CSA is also the only legislative act considered
that does not involve some operational considerations and requirements relating
to operations.

All laws rely on the conventional probabilistic understanding of the underlying
risk concepts. This observation could provide also a motivation for the public
authorities and stakeholders involved to evaluate whether non-probabilistic risk
concepts could enhance and improve existing risk analysis frameworks and
standards. Such an evaluation might be particularly worthwhile regarding other
risks than the man-made ones. In this regard, only the CER directive considers
also risks emerging from natural disasters and related phenomena. The CER
directive is also the only legislative act considered that does not operate with
a system-centric viewpoint. Nor does it (yet) entail clear-cut technical
development.

That said, the allowance of implementing and delegating acts in many of the
laws, such as those briefly remarked in Subsection~\ref{subsec: cer art 12 and
  13} regarding the CER directive, imply that the taxonomy may not be entirely
static. If required, it should be updated in the future by interpreting what the
supplementary acts may involve regarding risks (B12). This point is important to
underline also more generally. New laws and changes to existing laws often cause
ripple effects to existing implementations and their
designs~\cite{Ruohonen25ESPREb}. A similar point applies to guidance from public
authorities.

Ultimately, compliance questions as well as ambiguities are resolved through
enforcement actions, whether by public authorities or by courts. Regarding the
ambiguities, many of the legislative acts contain both explicit exemptions and
different qualifying words reflecting constraints and unconstraints for those
needing to comply (RQ.3). The CRA's exclusion of certain products is a good
example about explicit exemptions. Also qualifying words, such as
``\textit{appropriate}'' and ``\textit{proportionate}'', frequently appear in
most of the legislative acts analyzed. These hint that compliance is not a
clear-cut phenomenon but requires interpretation and different balancing
considerations. The GDPR is a good example. The wording in Articles 25 and 32
about the ``\textit{state-of-the-art}'' can be seen as a constraint for
technical design and implementation. Yet, it is balanced with many qualifying
words that represent relaxing unconstraints, including but not limited to
``\textit{costs}''.

\subsection{Takeaways}

The primary and perhaps obvious takeaway for practitioners would be that risks
should be taken into account when seeking to comply with the EU's core cyber
security laws. In fact, most of the laws discussed oblige to carry out risk
analyses for different purposes. These obligations lead to the second takeaway:
when needing to comply with multiple cyber security laws, it seems sensible to
recommend seeking synergies for compliance. Such synergies also underline the
term holistic~\cite{Haelterman22}. If a network-connected product processes
personal data, for instance, it seems sensible to consider the security
requirements of the GDPR in combination with the essential requirements of the
CRA. If further operating in a critical infrastructure sector or developing
products for such a sector, compliance burden could be likely reduced by
considering a single risk analysis covering the CRA's, CER's, and NIS2's
overlapping requirements. The remaining requirements could be than addressed on
case-by-case~basis.

Another example would be the CRA's essential cyber security requirements whose
design and implementation must be done based on a risk analysis. A sensible
approach might be to first evaluate whether all of them are applicable and
relevant, including with respect to the CIA triad, keeping in mind the CRA's
definition of a risk in mind. If a clear gap is present, a given requirement
should be likely implemented. Given the NIS2's emphasis on business continuity,
it might make sense to also consider economic and business aspects. That is:
when considering a distinction between value-adding requirements and residual
requirements causing negative business value~\cite{Fridgen14}, the CRA's notions
about an impact's magnitudes and losses or disruptions could be evaluated also
in business terms. A holistic approach is needed also when seeking compliance
with the GDPR because impacts should be evaluated against the rights and
freedoms of natural persons.

Given the numerous overlaps between different laws that were pointed out by
stakeholders during the CRA's policy consultation~\cite{Ruohonen25Bus}, further
research is also needed to continue the mapping studies already
done~\cite{Ruohonen25ESPREb} for better understanding overlapping legal
requirements. There also further overlaps present in terms of the risk
concepts. For instance, the EU's Digital Services Act~\cite{EU22c}, or DSA in
short, also uses a concept of systemic risks but in a very different context
than what was noted in Subsection~\ref{subsec: cra art 14, 15, 16, 17, 19, and
  20}. The AI regulation would be another good example.

However, these points about synergies and overlaps require understanding also
about acceptable and residual risks. These concepts too require further research
also in the context of the EU's cyber security laws. Given the NIS2's
all-hazards approach, it also remains unclear how well existing frameworks and
standards support encompassing risk analyses, including in relation to cascading
effects. Different propagated risk analysis methods~\cite{Waqdan25} might help
at modeling also larger hazards, disasters, and crises. A further related area
worth researching is related to the points raised about product-oriented
companies who may not have a clear understanding about a context in which their
products are used. In this regard, theoretical but adaptable scenarios, reusable
patterns and templates for scenarios and risk analysis matrices, better modeling
tools, and related topics might well be a research area with practical~value now
and in the future.

\balance
\bibliographystyle{apalike}
%\bibliography{eurisks}

\end{document}